\documentclass[graybox]{svmult}
\usepackage{mathptmx} 
\usepackage{helvet} 
\usepackage{courier} 
\usepackage{graphicx} 
\usepackage{url}
\makeindex 

\begin{document}

\title*{Towards DNS of the Ultimate Regime of Rayleigh--B\'enard Convection}
\author{Richard J.A.M. Stevens \and Detlef Lohse \and Roberto Verzicco}
\institute{R.J.A.M. Stevens \and D. Lohse\and R. Verzicco \at PoF, UTwente, \email{{r.j.a.m.stevens,d.lohse,r.verzicco}@utwente.nl}
\and R. Verzicco \at DII, Uniroma2 \email{verzicco@uniroma2.it}}
\maketitle

\section{Introduction}
\label{author_sec:1} 
Heat transfer mediated by a fluid is omnipresent in Nature as well as in technical applications and it is always among the fundamental mechanisms of the phenomena. The performance of modern computer processors has reached a plateau owing to the inadequacy of the fluid based cooling systems to get rid of the heat flux which increases with the operating frequency \cite{Chu}. On much larger spatial scales, circulations in the atmosphere and oceans are driven by temperature differences whose strength is key for the evolution of the weather and the stability of regional and global climate \cite{Egan}.

\begin{figure}[htb]
\sidecaption
\includegraphics[width=70mm]{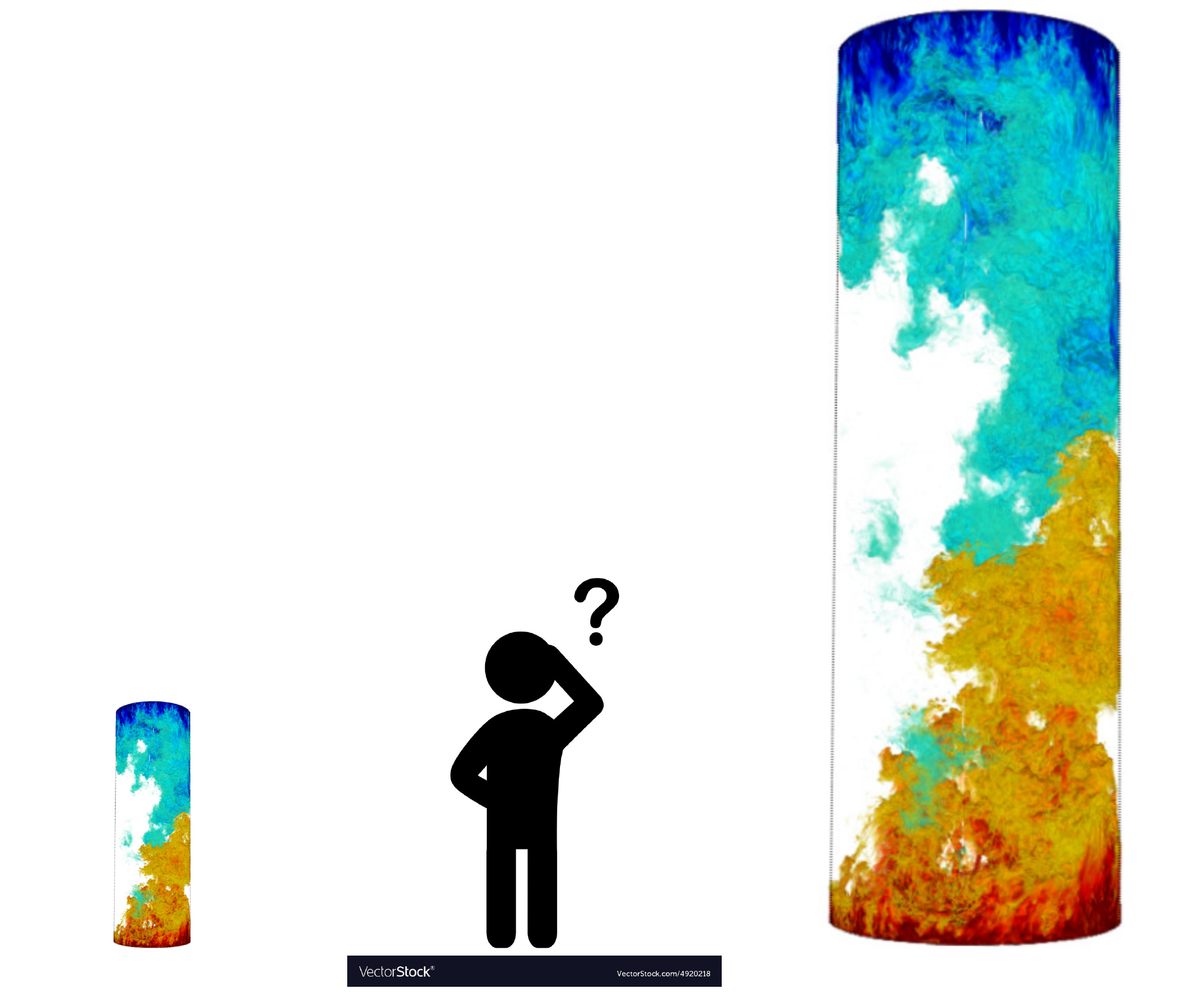} \hfill
\caption{Cartoon of the scaling problem for a Rayleigh--B\'enard flow: a model system of size $h_m$ has to be operated in dynamic similarity with a real system of size $h$. \label{author_fig:0}}
\end{figure}

The core of the problem, which is referred to as {\it natural convection}, is relatively simple since it reduces to determining the strength of the heat flux crossing the system for given flow conditions. Unfortunately, the governing equations (Navier--Stokes) are complex and non--linear, thus preventing the possibility to obtain analytical solutions. On the other hand laboratory experiments, aimed at tackling these phenomena, have to cope with the issue of how to make a setup of size $h_m = {\cal O}$(cm--m) dynamically similar to a system of $h={\cal O}$(m--Km) (figure \ref{author_fig:0}). Indeed, this upscaling problem is common to many experiments in hydrodynamics; in thermal convection however it is exacerbated since, as we will see in the next section, the most relevant governing parameter, the Rayleigh number ($Ra$), depends on the {\it third} power of the leading spatial scale. This implies that in real applications $Ra$ easily attains huge values while it hardly hits its low--end in laboratory experiments. 

Numerical simulations are subjected to similar limitations because of the spatial and temporal resolution requirements that become more severe as the Rayleigh number increases \cite{Shish10}. Only recently the former have become a viable alternative to experiments thanks to the continuously growing power of supercomputers.

Indeed, if the dynamics of the system could be expressed by power laws of the form $\approx ARa^\beta$, experiments and numerical simulations, performed at moderate values of the driving parameters, could be scaled up to determine the response of the real systems at extreme driving values. Unfortunately, this strategy works only assuming that the coefficients of the power law ($A$ and $\beta$) remain constant for every $Ra$ and this could not be the case for thermal convection \cite{GL}. More in detail, Malkus \cite{Malkus} and Priestley \cite{Priestley}, conjectured that all the mean temperature profile variations occur within the thermal boundary layers at the heated plates while the mean temperature in the bulk of the flow is essentially constant. Assuming also that the thermal boundary layers are far enough to evolve independently, one immediately obtains $\beta= 1/3$. A few years later, however, Kraichnan \cite{Kraichnan} noticed that as $Ra$ increases, also the flow strengthens and the viscous boundary layers eventually must become turbulent. In this case, referred to as {\it ultimate regime}, velocity profiles are logarithmic with the wall normal distance, and it results $\beta=1/2$ (times logarithmic corrections) which yields huge differences with respect to the previous theory.

This last observation and the fact that most of the practical applications evolve in the range of very high Rayleigh numbers motivate the effort to study turbulent thermal convection in the ultimate regime even if it requires the solution of formidable difficulties that we will detail in the next section.

\section{A trap problem}
\label{author_sec:2} One of the most appealing features of thermal convection is that the essence of the phenomenon can be reduced to a very simple model problem in which a fluid layer of thickness $h$, kinematic viscosity $\nu$ and thermal diffusivity $\kappa$ is heated from below and cooled from above with a temperature difference $T_h-T_c=\Delta$. The temperature field, in a constant gravity field $g$ produces a flow motion via the thermal expansion, parameterized by a constant coefficient $\alpha$: this is the Rayleigh--B\'enard flow. The heat flux $\dot Q$ between the plates will be a function of the form $\dot Q = f (h,\nu,\kappa,g,\alpha,\Delta)$ which involves $N=7$ different quantities whose minimum number of independent dimensions is $K=4$. The Buckingham $\Pi$--theorem assures that, in non--dimensional form, the above relation is equivalent to one with only $N-K=3$ parameters that can be written as $Nu=F(Ra,Pr)$ being $Ra= g\alpha \Delta h^3/(\nu\kappa)$, $Pr=\nu/\kappa$ and $Nu=\dot Q/\dot Q_{diff}$ with $Q_{diff}$ the diffusive heat flux through the fluid in absence of motion.

The above relation looks very attractive since it depends only on two independent parameters $Ra$ and $Pr$, the latter being determined solely by the fluid properties. Many efforts have been made to study the function $F(\cdot)$ through laboratory experiments and, in the last two decades, also by numerical simulations as a viable alternative. Unfortunately, the practical realization of the RB convection is substantially different from the theoretical problem and many additional details come into play.

The first point is that while in the ideal flow the fluid layer is laterally unbounded, in real experiments, by necessity, it must be somehow confined thus introducing a second length $d$ or equivalently an additional parameter $\Gamma = d/h$, the aspect--ratio. The parameters become more than one if the tank does not have a cylindrical or square cross section. 

\begin{figure}[t]
\centering
\begin{minipage}{10cm}
\leftfigure{\includegraphics[width=45mm]{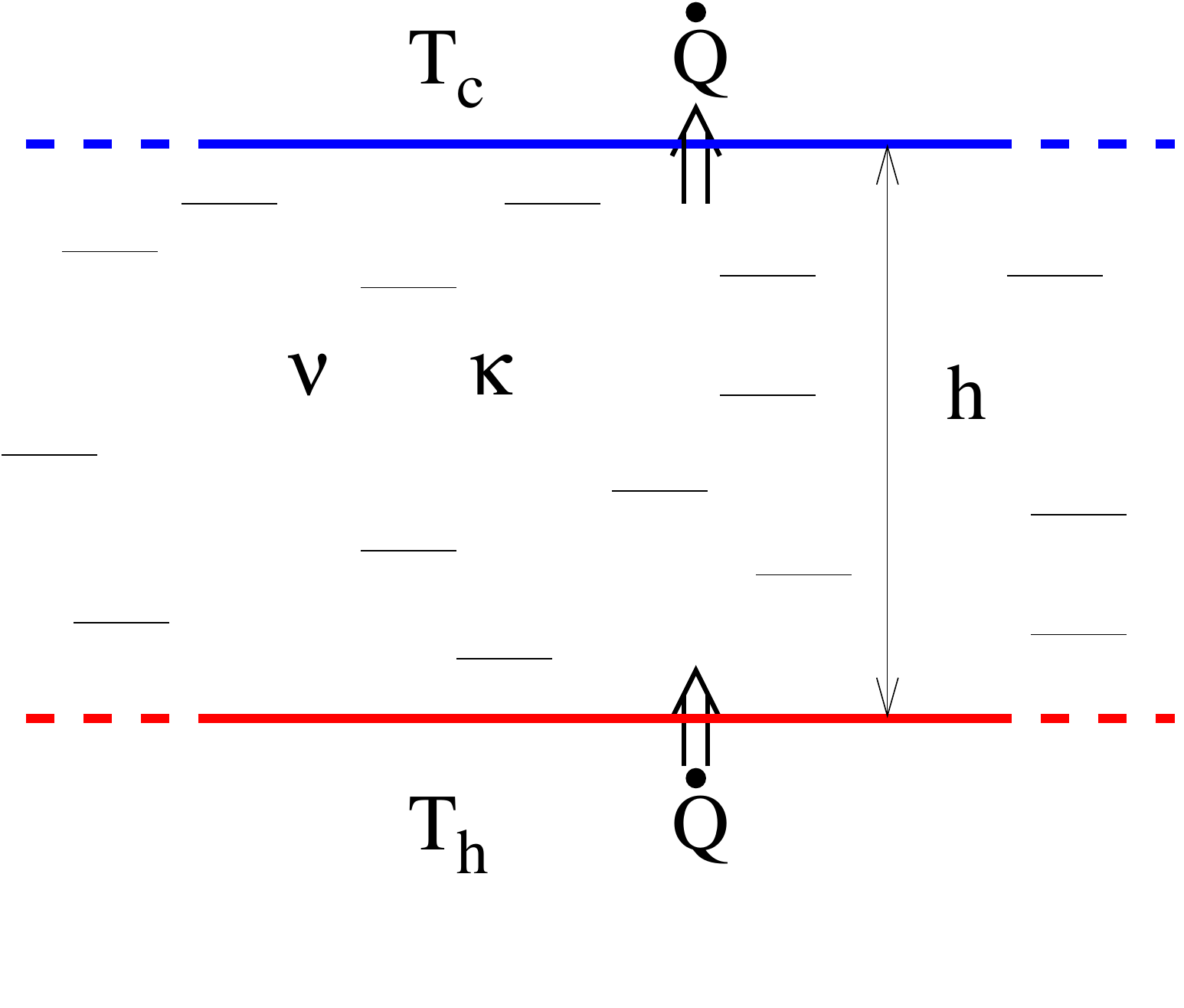}} \hfill
\rightfigure{\includegraphics[width=45mm]{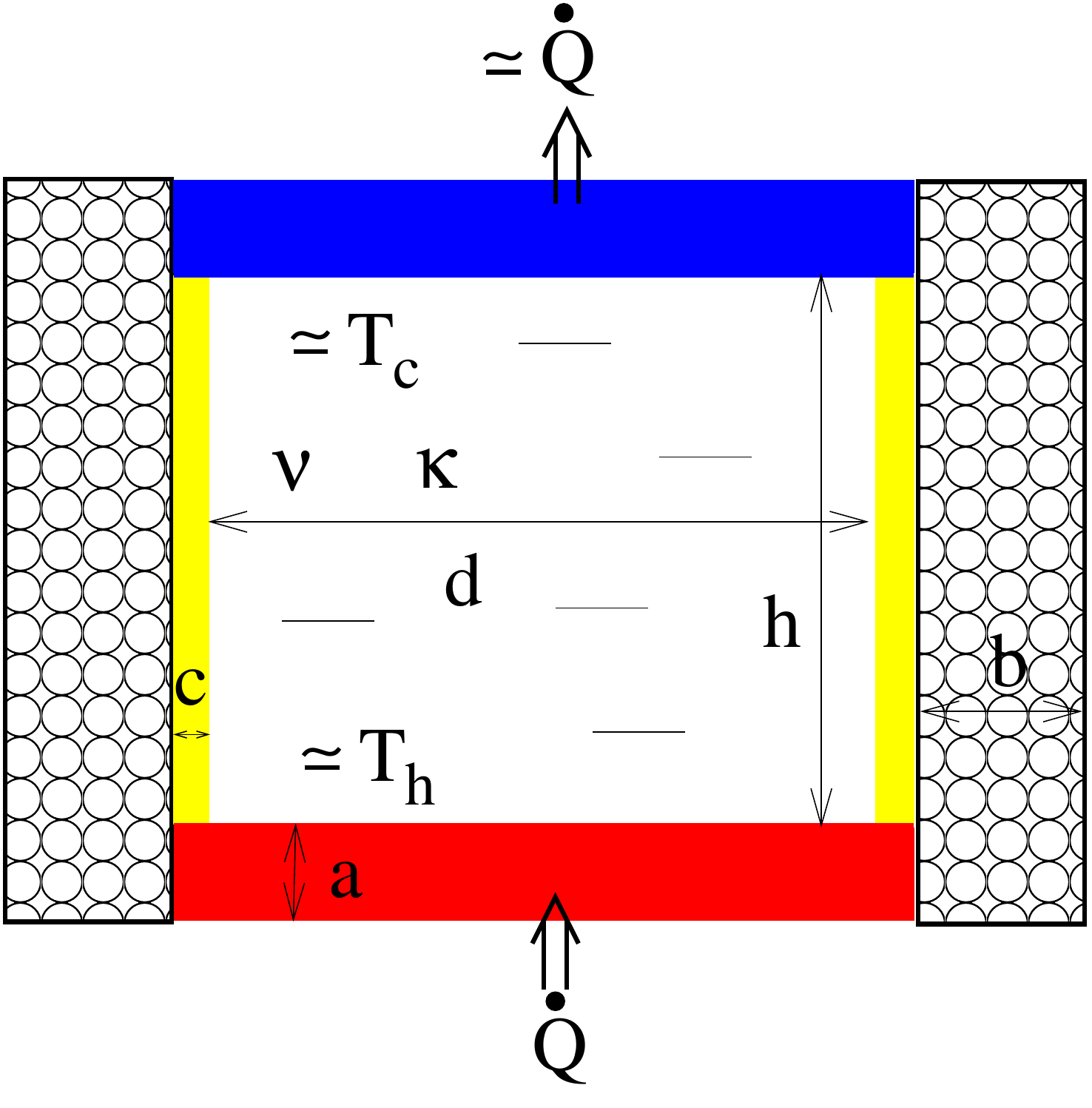}}
\leftcaption{Cartoon of the ideal Rayleigh--B\'enard flow. \label{author_fig:1a}}
\rightcaption{Simplified sketch of a real experimental set--up. Outside of the cell only an insulating layer of foam has been reported while all the details about heating and cooling systems and thermal shields have been neglected. \label{author_fig:1b}}
\end{minipage}
\end{figure}

Several additional variables are introduced by the physical realization of the thermal boundary conditions; in fact, in the ideal problem, all the heat entering the fluid through the lower isothermal hot plate leaves the fluid only crossing the upper cold plate, without heat leakage across the side boundaries. In the real flow, the isothermal surfaces are obtained by thick metal plates ($a={\cal O}(1$--$5)$\ cm) of high thermal diffusivity $\kappa_{pl}$ (copper or aluminum) that provide stable temperature values at the fluid interface for every flow condition. The sidewall, in contrast, should minimize the heat transfer and it is therefore made of low thermal diffusivity $\kappa_{sw}$ materials (steel or Plexiglas) and with reduced thickness ($e={\cal O}(1$--$5)$\ mm). To further prevent parasite heat currents, insulation foam layers and even active thermal shields are installed outside of the cell, thus further increasing the number of input parameters. Within this scenario, the heat flux function looks like $ \dot Q = f' (h,\nu,\kappa,g,\alpha,\Delta,d,e,\kappa_{sw},a,\kappa_{pl},b,\kappa_{foam},...)$ whose variable counting is $N > 14$ while it results always $K=4$ thus implying that the non--dimensional counterpart ($Nu = F'(Ra,Pr,...)$) involves more than $N-K>10$ independent parameters.

\begin{figure}[htb]
\sidecaption
\includegraphics[width=75mm]{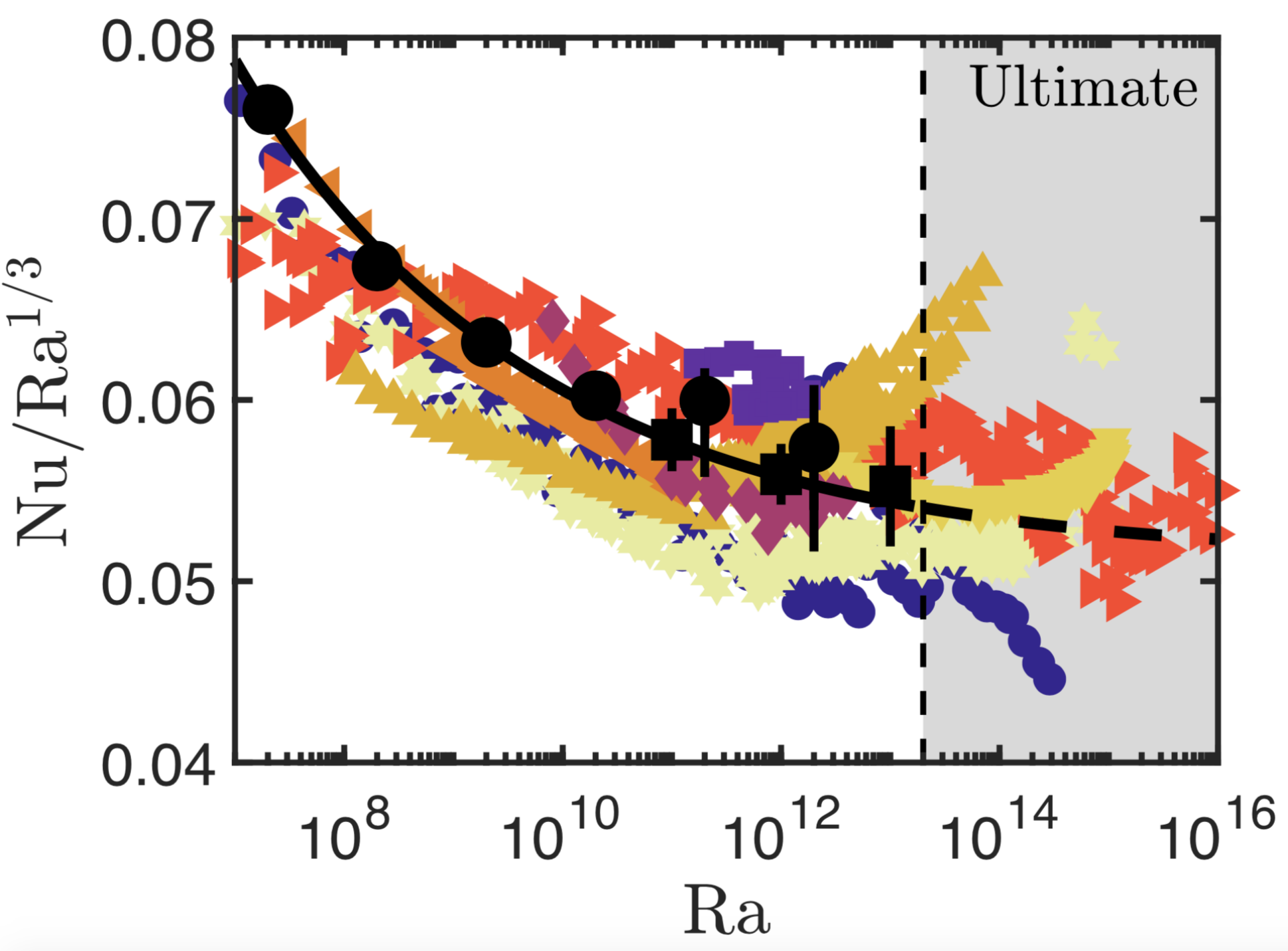} \hfill
\caption{Compilation of experimental and numerical data for the compensated Nusselt versus Rayleigh numbers:
the solid black line is the theory by \cite{GL},
blue bullets are the experiments by \cite{Castaing},
purple squares \cite{Fleisher},
dark red diamonds \cite{Chaumat},
red right triangles \cite{Niemela},
orange left triangles \cite{Ahlers},
yellow up triangles \cite{Roche},
yellow down triangles \cite{He},
yellow stars \cite{Urban},
black bullets, numerical simulations by \cite{Stevens11},
black squares, preliminary numerical simulations by Stevens (2019),
({\it Personal Communication}).
\label{author_fig:2}}
\end{figure}

The hope is that when the function $F'(\cdot)$ is explored, experimentally or numerically, most of the parameters introduced by the experimental technicalities do not affect significantly the phenomena and the relevant variables reduce to a tractable number. Unfortunately, some of the recent, and not so recent, experiments have shown that it is not always the case since dynamically equivalent flows do not yield identical results. In figure \ref{author_fig:2} we report a collection of Nusselt numbers taken from different sources showing some disagreement both at the low-- and high--end of $Ra$. While the former differences have been attributed mainly to the heat leakage through the sidewall \cite{Roch,Ahl,Ver1,Stev1} and some of the latter to the non--perfect thermal sources \cite{Ver2}, most of the discrepancies are still unexplained and they are the subject of intense investigation by many research groups worldwide. The situation is even more complex if one does not restrict the analysis only to the Nusselt number since higher order statistics show higher sensitivity to external perturbations.

In this context, numerical experiments can be particularly helpful since they can be used as ideal tests to isolate the different perturbations of the basic problem and assess their effect. However, performing direct numerical simulation of Rayleigh-B\'enard convection implies several non obvious choices and requires huge computational resources that, for the parameter range of the ultimate regime, are not fully available yet. In the following we will present estimates of the computational costs and discuss some of the possible open choices. 

The first relevant point is whether the simulation should be aimed at the ideal RB flow or rather has to mimic a laboratory experiment. In the first case the computational domain should be laterally unbounded which can be approximated by periodic boundary conditions applied to a rectangular domain of horizontal size $d$; this configuration results in a horizontally homogeneous flow that benefits from easy and efficient uniform spatial discretizations and fast converging statistics. On the other hand, in a real set--up, the boundedness of the fluid layer generally results in a smaller fluid volume although the kinematic boundary layer at the sidewall has to be resolved by additional gridpoints. 

\begin{figure}[htb]
\sidecaption
\includegraphics[width=75mm]{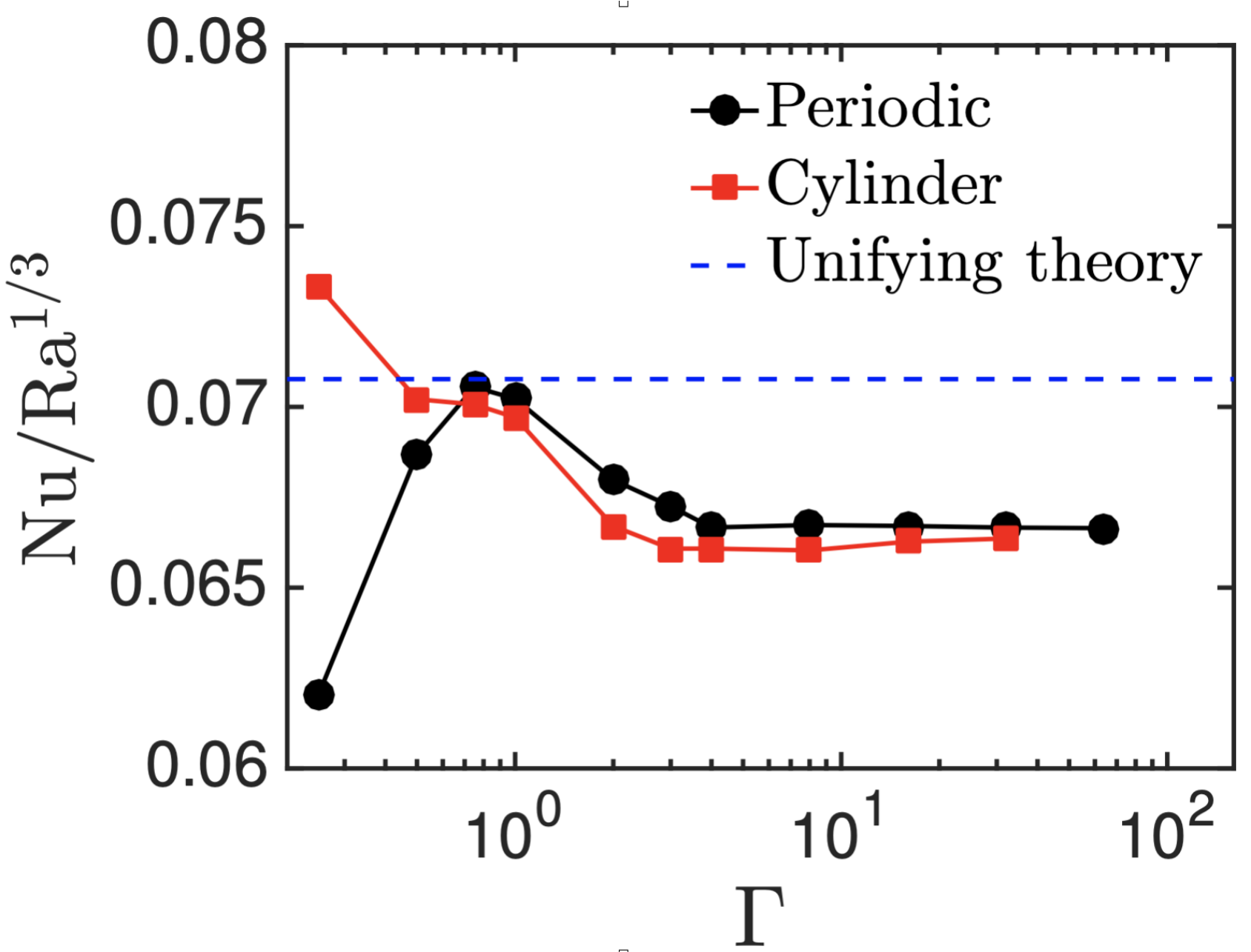} \hfill
\caption{Nusselt number versus domain aspect--ratio $\Gamma=d/h$ at $Ra=10^8$ and $Pr=1$. The dashed line for the `unifying theory' \cite{GL} is computed from experimental data at $\Gamma=1$.
\label{author_fig:3}}
\end{figure}

Which of the two configurations is more advantageous is not obvious since the computational efficiency of the fully bounded flow solvers is smaller than those with homogeneous directions but the former setup involves smaller flow volumes.

Besides computational considerations there are also physical issues since the presence of lateral boundaries, and even the shape of the container affect the flow dynamics \cite{Daya}. In figure \ref{author_fig:3} we report the dependence of the Nusselt number on the domain aspect--ratio $\Gamma$ for rectangular and cylindrical geometries. It is immediately evident that, for slender domains $Nu$ depends on $\Gamma$ in the opposite way and only for $\Gamma \approx 1$ the two geometries yield similar values while for $\Gamma \geq 4$ the Nusselt number converges to the asymptotic value for unbounded domains. All the laboratory experiments reported in figure \ref{author_fig:2} have been performed in low aspect ratio cylindrical cells and accordingly also the numerical simulations have been run in a cylinder at $\Gamma=0.5$. In contrast, if a horizontally homogeneous flow has to be simulated, in order to get rid of the numerical confinement effect, it must be computed on domains at least of $\Gamma=4$, even if spectra and higher order statistics indicate that $\Gamma=8$ or $\Gamma=16$ is needed to eliminate confinement effects \cite{Blass}. 

In the following we give an estimate of the computational resources needed for direct numerical simulation of turbulent RB convection in rectangular and cylindrical geometries by evaluating the number of nodes contained in the relative mesh. The basic assumption is that the flow can be divided into bulk and boundary layer regions, the former discretized by a mesh of the same size as the smallest between the Kolmogorov and Batchelor scales and the latter with the resolution criteria suggested by \cite{Shish10}. We further assume that the rectangular box has a size $d\times d\times h$ discretized in Cartesian coordinates while the cylinder has a diameter $d$ and a height $h$ discretized in polar coordinates.

For the ease of discussion we will restrict to $Pr=1$ keeping in mind that as the Prandtl number deviates substantially from unity the simulation becomes even more demanding either because the velocity field develops finer scales than the temperature ($Pr \ll 1$) or vice versa ($Pr\gg 1$).

For the mean Kolmogorov scale $\eta$ we can easily write $\eta/h\approx (RaNu)^{1/4}$ that with a fit $Nu = A Ra^\beta$ ($A\simeq 0.05$ and $\beta = 1/3$ from the high end of $Ra$ in figure \ref{author_fig:2}) yields a number of nodes per unit length in the bulk $N_{bu} = 0.473Ra^{1/3}$. For the resolution of each boundary layer we rely on the correlation derived by \cite{Shish10} which suggest a number of nodes $N_{bl} \approx 0.35 Ra^{0.15}$. Within these figures, the total number of nodes for the rectangular domain reads $N_{Car} = \Gamma^2 (0.105Ra + 0.156Ra^{0.816})$.

We proceed along the same lines for the cylindrical domain keeping in mind that there is an extra boundary layer at the sidewall and that the polar coordinates have azimuthal isolines that diverge radially. Therefore the resolution requirements in this direction are dictated by the location farthest from the symmetry axis. Using the same correlations as above we obtain $N_{Cyl} = 0.5\pi \Gamma^2 (0.105Ra + 0.156Ra^{0.816}) + \pi \Gamma (0.223Ra^{0.816}+0.116Ra^{0.633})$.

It is worth mentioning that these expressions have been obtained by simplifying assumptions therefore their results should be taken as coarse estimates and not as precise measures. For example, $A$ and $\beta$ have been assumed constant and equal to the high--end $Ra$ values of figure \ref{author_fig:2} and we have used $h -2 \delta_{bl} \approx h$ (with $\delta_{bl}$ the boundary layer thickness): all these positions concur to an overestimate of the number of nodes. On the other hand, the correlation $N_{bl} \approx 0.35 Ra^{0.15}$ of \cite{Shish10} was obtained for a Prandtl--Blasius laminar boundary layer that is expected to underestimates the resolution when the ultimate regime sets in and the boundary layers transition to turbulence. At the transitional Rayleigh number, the above factors might compensate each other and the estimates could give reasonable numbers.

A comparison of the two expressions immediately shows that the leading order term increases at the same rate with $Ra$ and $\Gamma$ although the cylindrical mesh has asymptotically $60\%$ more nodes than the Cartesian counterpart. This is true although, for the same aspect--ratio, the latter has a volume ($\Gamma^2h^3$) which is more than $20\%$ bigger than the former ($\pi \Gamma^2h^3/4$).

\begin{figure}[htb]
\sidecaption
\includegraphics[width=75mm]{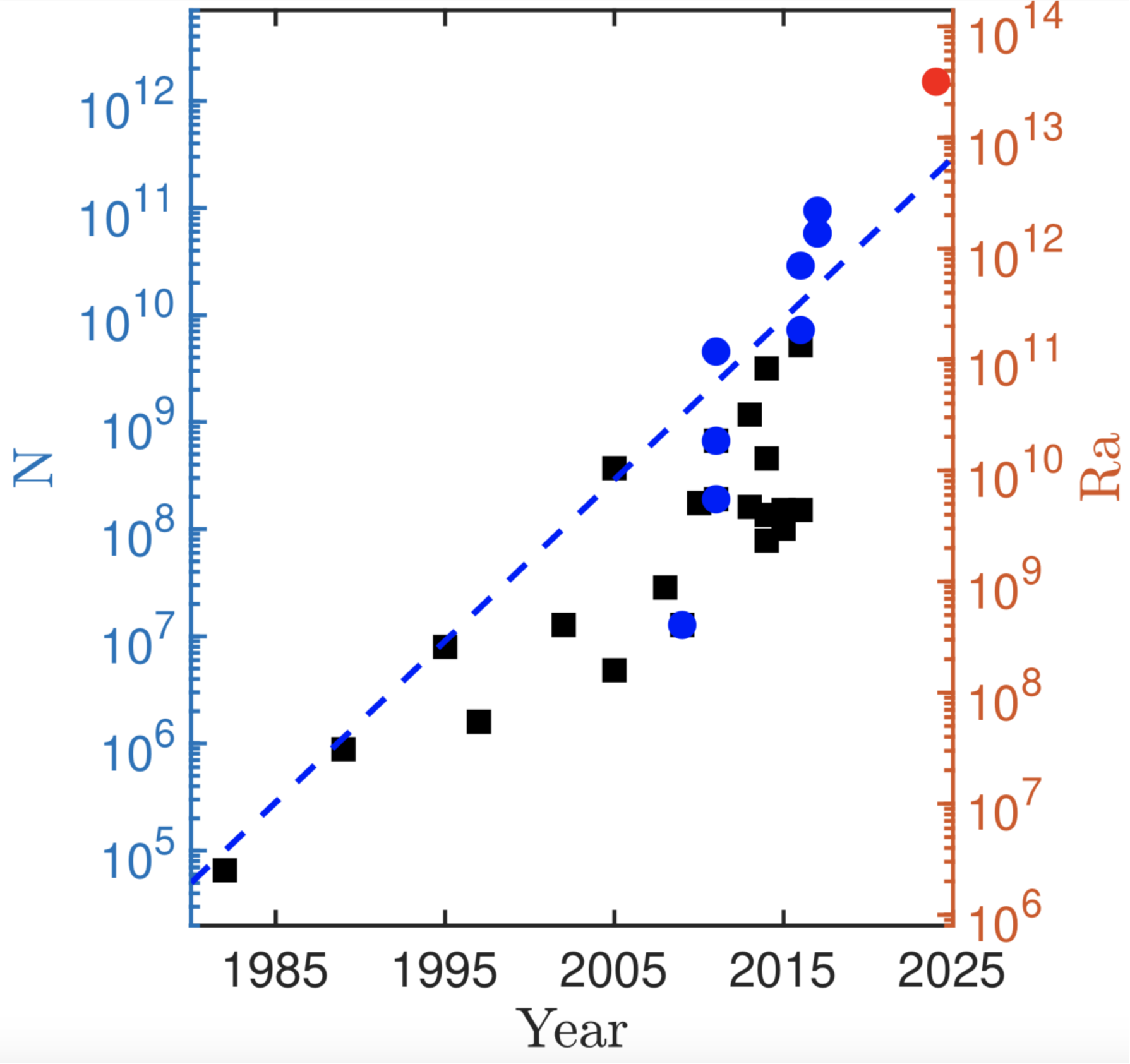} \hfill
\caption{Number of nodes $N$ (in a cylindrical cell of $\Gamma=0.5$) and achievable flow Rayleigh number $Ra$ versus the years for direct numerical simulations of Rayleigh--B\'enard convection. Black squares for various data from the literature, blue bullets for simulations from our research group, big red bullet final goal for the ultimate regime simulation.
\label{author_fig:4}}
\end{figure}

If now we focus on the onset of the ultimate regime we have to determine the critical Rayleigh number at which the boundary layer undertakes the transition to the turbulent
state. This is triggered by the large scales of convection that sweep the plates by the induced winds; according to Reference \cite{Landau} the boundary layer transition occurs for a shear Reynolds number of $Re_S \approx 420$ that Grossmann \& Lohse \cite{GL} have estimated to happen around $Ra_C \approx 10^{14}$. 

In a rectangular domain with $\Gamma=4$ this Rayleigh number implies a mesh with $N_{Car} > 10^{14}$ nodes that is clearly infeasible in the mid--term future. In a cylindrical cell of $\Gamma=0.5$, however, it results $N_{Cyl} \approx 10^{12}$ nodes that could be achieved within the next five years (see figure \ref{author_fig:4}). Indeed, we are already running simulations at $Ra=10^{13}$ at $\Gamma=0.5$ and even $Ra=10^{14}$ at $\Gamma=0.25$ with meshes of the order of $10^{11}$ nodes (R. Stevens, {\it Personal Communication}) although we expect to tackle the ultimate regime only by the `next generation' simulations. 

We wish to point out that if we compare the numbers coming from the present formulas with those currently used for the highest Rayleigh number simulations we find that the former produce a systematic overestimate of the required resolution. For example, for a cylindrical mesh of aspect--ratio $\Gamma = 0.5$ at $Ra=10^{13}$ our prediction yields a number of nodes $N_{Cyl} \approx 4\times 10^{11}$ while a simulation on a mesh $4608\times1400\times4480$ ($N_{Cyl} \approx 2.9\times 10^{10}$) yielded the same Nusselt number as another run on the finer grid $6144\times1536\times6144$ with $N_{Cyl} \approx 5.8\times 10^{10}$ nodes (R. Stevens, {\it Personal Communication}).

A possible explanation for this difference is that in our model we have assumed that in the bulk the mesh has to be as fine as the mean Kolmogorov scale $\eta$. However, looking at the dissipation spectra of turbulence \cite{Pope} one finds its peak around $10\eta$ thus implying that also a mesh of size $1.5$--$2\eta$ already resolves most of the dissipation. In three--dimensions this difference yields a factor $6.25$--$8$ less in the node counting that is about the mismatch between our prediction and the actual meshes.

For a while we have been working at improving the simulation code \cite{VO,VC,Stevens11} by more efficient implementations of the solution algorithms and of the parallelization strategies in order to reduce the time--to--solution. In addition we are also figuring out alternatives to achieve the ultimate regime in more affordable problems.

One possible way is to exploit the analogy between Rayleigh--B\'enard and Taylor--Couette (TC) flow \cite{Busse}. The latter is the flow developing in the gap between two coaxial cylinders rotating at different angular velocities and whose angular momentum flux across the cylinders behaves as the heat flux between the plates in a RB flow \cite{Bruno}. It turns out, however, that the mechanical forcing of the TC flow is more efficient in producing turbulent boundary layers than the thermal forcing of RB flows and the ultimate regime can be achieved for smaller values of the driving parameters that are affordable by numerical simulation \cite{Ostilla}. In Reference \cite{Zhu}, thanks to the presence of baffled cylinders, that disrupted the logarithmic part of the turbulent boundary layer profiles, it has been possible to get rid of the logarithmic correction and obtain a pure $1/2$ power law in the analogous of the $Nu$ versus $Ra$ relationship.

Another possibility is to simulate a two--dimensional RB flow that allows, already now, to tackle Rayleigh numbers $> 10^{14}$; indeed in Reference \cite{Zhu2} (and successive developments) simulations have been run up to $Ra \simeq 5\times 10^{14}$ with the appearance of a transition already for $Ra \geq 10^{13}$. 

\section{Closing remarks} \label{author_sec:3} 
In this contribution we have briefly introduced the problem of turbulent thermal convection with a particular look at its transition to the ultimate regime and the resolution requirements needed for the direct numerical simulation of this flow.

Leaving aside all the complications related to the spurious heat currents through the sidewall and the imperfect character of the thermal sources, already addressed in some of the referred papers, it appears that a preliminary fundamental question is whether the simulation should be aimed at replicating an experimental set--up with a lateral confinement or to mimic the truly Rayleigh--B\'enard flow that is virtually infinite in the horizontal directions.

We have shown that in the latter case a domain with aspect--ratio no smaller than $\Gamma =4$ is required and this implies, at the estimated critical Rayleigh number $Ra_C \approx 10^{14}$, a computational mesh with more than $10^{14}$ nodes that is not likely to be tractable within the next decade. On the other hand, although for a given $Ra$ and $\Gamma$ cylindrical, laterally confined geometries contain about $60\%$ more nodes than the rectangular `unbounded' domains, when restricted to the existing, slender cylinders of the laboratory experiments the number of nodes becomes more feasible. In particular, for $\Gamma=0.5$ and $Ra = 10^{14}$ the present estimate gives a mesh slightly larger that a trillion of nodes. Even if this number might look impressive, it is `only' one order of magnitude bigger than the current state--of--the--art simulations and, according to figure \ref{author_fig:4}, such meshes will become affordable within the next five years or so. It is also worthwhile mentioning that the present estimates assume a mesh in the bulk of the flow that is everywhere as fine as the mean Kolmogorov scale $\eta$ while actual grid refinement checks performed on Rayleigh--B\'enard turbulence have shown converged results already for meshes of size $2\eta$. This implies that in three--dimensional flows the actual mesh sizes can be about one order of magnitude smaller and this is fully confirmed by our ongoing simulations.

Needless to say, once the ultimate regime will have been hit by numerical simulations also in three--dimensions, a {\it terra incognita} will be entered. Turbulent boundary layers have more severe resolution requirements than the laminar counterparts and once the ballistic plumes of Kraichnan \cite{Kraichnan}, which can be thought of as pieces of detached thermal boundary layer, are shot into the bulk also the resolution of that flow region is likely to become more demanding. Clearly, attempting resolution estimates beyond the onset of the transition would be even more speculative than those of the present paper and only with the data of those simulations at hand, further reasonable projections can be made.

While waiting for adequate computational resources to tackle thermal convection in the ultimate regime we can nevertheless compute turbulent flows that exhibit similar dynamics or that can be reduced to a tractable size by simplifying assumptions. These include the Taylor--Couette flow, that can be rigorously shown to be analogous to Rayleigh-B\'enard convection, or two--dimensional thermal convection that already now can be simulated well beyond $Ra=10^{14}$ and has indeed shown evidence of transition to the ultimate state.\\

\textbf{Acknowledgements} This work is supported by the Twente Max-Planck Center and the ERC (European Research Council) Starting Grant no. 804283 UltimateRB. The authors gratefully acknowledge the Gauss Centre for Supercomputing e.V. ({\color{blue}\url{www.gauss-centre.eu}}) for funding this project by providing computing time on the GCS Supercomputer SuperMUC-NG at Leibniz Supercomputing Centre ({\color{blue}\url{www.lrz.de}}).


\begin{thebibliography}{99.}
\bibitem{Chu} Chu, R.C., Simons, R.E., Ellsworth, M.J., Schmidt, R.R., Cozzolino, V.:
Review of cooling technologies for computer products,
IEEE Trans. on Dev. and Mat. Reliab., \textbf{4}(4), 568--585, (2004).
\bibitem{Egan} Egan, P.J., Megan Mullin, M.: Recent improvement and projected worsening of weather in the United States, Nature, \textbf{532}, 357--360, (2016).
\bibitem{Shish10} Shishkina, O., Stevens, R.J.A.M., Grossmann, S., Lohse, D.: 
Boundary layer structure in turbulent thermal convection and its consequences 
for the required numerical resolution.
 New J. Phys., \textbf{12}, 075022, (2010).
\bibitem{GL} Grossmann, S., Lohse, D.: Scaling in thermal convection: a unifying theory.
 J. Fluid Mech., \textbf{407}, 27--56, (2000).
\bibitem{Malkus} Malkus, M.V.R.: Heat transport and spectrum of thermal turbulence. Proc. R. Soc. London, Ser. A,
\textbf{225}, 169, (1954).
\bibitem{Priestley} Priestley, C.H.B.: Turbulent Transfer in the Lower Atmosphere, University of Chicago Press, Chicago,
(1959).
\bibitem{Kraichnan} Kraichnan, R.H.: Turbulent Thermal Convection at Arbitrary Prandtl Number. Phys. of Fluids,
\textbf{5}(11), 1374, (1962).
\bibitem{Castaing} Castaing, B., Gunaratne, G., Heslot, F., Kadanoff., L., Libchaber, A., Thomae, S., Wu, X. Z.,
Zaleski, S. \& Zanetti, G..: Scaling of hard thermal turbulence in Rayleigh-B\'enard convection.
 J. Fluid Mech., \textbf{204}, 1--30, (1989).
\bibitem{Fleisher} Fleisher, A.S., Goldstein, R.J.: High Rayleigh number convection of pressurized 
gases in a horizontal enclosure. J. Fluid Mech., \textbf{469}, 1--12, (2002).
\bibitem{Chaumat} Chaumat, S., Castaing, B., Chilla, F.: Rayleigh--B\'enard cells: influence of plate properties. In Advances in Turbulence IX (ed. Castro, I. P., Hancock, P. E. and Thomas, T. G.). International Center for Numerical Methods in Engineering, CIMNE.
\bibitem{Niemela} Niemela, J., Skrbek, L., Sreenivasan, K.R., Donnelly, R.: Turbulent convection at 
very high Rayleigh numbers. Nature, \textbf{404}, 837–840, (2000).
\bibitem{Ahlers} Ahlers, G., He, X., Funfshilling, D., Bodenshatz, E.: Heat transport by turbulent 
Rayleigh–B\'enard convection for $Pr\simeq 0.8$ and $3\times10^{12}\leq Ra \leq 10^{15}$:
aspect ratio $\Gamma = 0.50$. New J. Phys., \textbf{14}, 103012, (2012).
\bibitem{Roche} Roche, P.E., Gauthier, F., Kaiser, R., Salort, J.: On the triggering of the 
ultimate regime of convection. New J. Phys., \textbf{12}, 085014, (2012).
\bibitem{He} He, X., Funfshilling, D., Bodenshatz, E., Ahlers, G.: Heat transport by turbulent 
Rayleigh–B\'enard convection for $Pr\simeq 0.8$ and $4\times10^{11}\leq Ra \leq 2\times 10^{14}$:
aspect ratio $\Gamma = 1.00$. New J. Phys., \textbf{14}, 103012, (2012).
\bibitem{Urban} Urban, P., Musilovs\'a, V., Skrbek. L.: Efficiency of heat transfer in turbulent 
Rayleigh--B\'enard convection, Phys. Rev. Lett., \textbf{107}, 014302, (2011).
\bibitem{Stevens11} Stevens, R.J.A.M., Lohse, D., Verzicco, R.: Prandtl and Rayleigh number 
dependence of heat transport in high Rayleigh number thermal convection,
 J. of Fluid Mech., \textbf{688}, 31--43, (2011).
\bibitem{Roch} Roche, P. E., Castaing, B., Chabaud, B., Hebral, B., Sommeria, J.: Side wall effects in Rayleigh--B\'enard experiments. Eur. Phys. J. B, \textbf{24}, 405--408, (2001).
\bibitem{Ahl} Ahlers, G.: Effect of sidewall conductance on heat-transport measurements for turbulent Rayleigh--B\'enard convection. Phys. Rev. E, \textbf{63}, 015303, (2000).
\bibitem{Ver1} Verzicco, R.: Sidewall finite conductivity effects in confined turbulent thermal convection. J. Fluid Mech., \textbf{473}, 201--210, (2002).
\bibitem{Stev1} Stevens, R.J.A.M., Lohse, D., Verzicco, R.: Sidewall effects in Rayleigh--B\'enard convection, J. Fluid Mech., \textbf{741}, 1--27, (2014).
\bibitem{Ver2} Verzicco, R.: Effect of non perfect thermal sources in turbulent thermal
convection, Phys. of Fluids, \textbf{16}, 1965, (2004).
\bibitem{Daya} Daya, Z. A., Ecke, R. E.: Does turbulent convection feel the 
shape of the container?. Phys. Rev. Lett., \textbf{89}, 4501, U81–U83, (2001).
\bibitem{Blass} Stevens, R.J.A.M., Blass, A., Zhu, X., Verzicco, R., Lohse, D.: Turbulent thermal superstructures in Rayleigh-B\'enard convection, Phys. Review Fluids, \textbf{3}(4), 041501, (2018).
\bibitem{Landau} Landau, L.D., Lifshitz, E.M.: Fluid Mechanics. Pergamon Press, Oxford (1984)
\bibitem{Pope} Pope, S.B.: Turbulent Flows. Cambridge University Press, Cambridge (2000)
\bibitem{VO} Verzicco, R., Orlandi P.: A finite--difference scheme for three dimensional incompressible flows in cylindrical coordinates, J. Comp. Phys., \textbf{123}, 402, (1996).
\bibitem{VC} Verzicco, R., Camussi, R.: Numerical experiments on strongly turbulent thermal convection in a slender cylindrical cell, J. Fluid Mech., \textbf{477}, 19--49, (2003).
\bibitem{Busse} Busse, F.H.: Viewpoint: The Twins of Turbulence Research, Physics, \textbf{5}, 4, (2012).
\bibitem{Bruno} Eckhardt, B., Grossmann, S., Lohse, D.: Torque scaling in turbulent Taylor--Couette flow between independently rotating cylinders. J. Fluid Mech., \textbf{581}, 221--250, (2007).
\bibitem{Ostilla} Ostilla--Monico, R., van der Poel, E., Verzicco, R. Grossmann, S., Sun, C., Lohse, D.: Exploring the phase diagram in the fully turbulent Taylor--Couette flow. J. Fluid Mech., \textbf{761}, --26, (2014).
\bibitem{Zhu} Zhu, X., Verschoof, R.A., Bakhuis, D., Huisman, S.G., Verzicco, R., Sun, C., Lohse, D.: Wall roughness induces asymptotic ultimate turbulence, Nature Phys., \textbf{14}(4), 417--423, (2018).
\bibitem{Zhu2} Zhu, X., Mathai, V., Stevens, R.J.A.M., Verzicco, R., Lohse, D.: Transition to the Ultimate Regime in Two-Dimensional Rayleigh-B\'enard Convection. Phys. Review Lett., \textbf{120}(14), 144502, (2018).

\end{thebibliography}
\end{document}